# The Cloud Technologies and Augmented Reality: the Prospects of Use

Maiia V. Popel[0000-0002-8087-962X] and Mariya P. Shyshkina[0000-0001-5569-2700]

Institute of Information Technologies and Learning Tools of NAES of Ukraine,
9, M. Berlynskoho St., Kyiv, 04060, Ukraine
`popelmaya@gmail.com, shyshkina@iitlt.gov.ua`

**Abstract.** The article discusses the prospects of the augmented reality using as a component of a cloud-based environment. The *research goals* are the next: to explore the possibility of the augmented reality using with the involvement of the cloud-based environment components. The *research objectives* are the next: to consider the notion of augmented reality; to analyze the experience the augmented reality using within the cloud environment / system; to outline the prospects of the augmented reality using in educational institutions; to consider the technical conditions of the augmented reality use. The *object of research* is: the educational process in educational institutions of Ukraine of different levels of accreditation. The *subject of research* is: the educational process in a cloud-based environment in educational institutions of Ukraine. The *research methods* used are the next: analysis of scientific publications, observations. The *results of the research* are the next: on the basis of the analysis of scientific works, it has been established that the experience of the augmented reality using in the systems based on cloud technologies already exists. However, the success of such a combination has not yet been proven. Currently, laboratory tests are known, while the experiment was not carried out under natural conditions in control and experimental groups. It is revealed that the attraction of the augmented reality for the educators requires the development of new methodologies, didactic materials, updating and updating of the curriculum. The *main conclusions and recommendations*: the main principles of augmented reality use in the learning process are: designing of the environment that is flexible enough, attention should be paid to the teaching and didactic issues; adjusting the educational content for mastering the material provided by the curriculum; the research methods that can be used in training along with the elements of augmented reality are to be elaborated; development of adaptive materials; training of teachers, which will include augmented reality in educational practice.

**Keywords:** augmented reality, cloud computing, cloud technologies, cloud-based environment, component of the cloud-oriented environment.

Cloud services have long been attracting the brisk interest of scholars in all areas. The educational sciences have not become an exception. However, there was a problem of lack of the learning methods of these technologies use; the methods of designing of the specific environments for higher and secondary educational institutions of Ukraine, adapted to the needs of teachers and students, teachers and pupils. A plenty of works (e.g., [6; 7; 8; 9; 13; 16]) were devoted to the solving of a number of problems concerning the cloud services educational implementation. The number of universities and institutions in Ukraine that use cloud services not only to save money, but to increase the efficiency and convenience of educational services providing is constantly increasing.

However, the classical methods and techniques of pedagogical science suffer changes related to the introduction of new learning tools. Pupils and students are interested in new technologies based on virtual and augmented reality. In addition, there is a point of view that the use of augmented reality in teaching activities will simplify the learning process, provide an opportunity to save money on demonstration materials and does not require the purchase of additional equipment (for example, the smartphone has almost every student and pupil).

According to the Michael A. Dougherty, Samuel A. Mann, Matthew L. Bronder, Joseph Bertolami, and Robert M. Craig [4] augmented reality – is a combination of real-world data and computer data to create a unified user environment. Real-world data can be collected using an appropriate data collection method such as by means of a camera or a chip of the phone. These data can be processed and merged with the data developed by the computer to create the user's environment. One of the most common forms of augmented reality is the use of video images taken and processed by a video camera, with the increased graphics from a PC or other image.

That is, for most institutions that have deployed the cloud environment, it would be possible to include into it the components of augmented reality, or link it with existing didactic material. Installing the module for learning using augmented reality in the cloud is to provide a more efficient exchange of information between teachers and developer services, students or pupils. By providing access to the augmented reality module through the cloud environments, you can solve the problem of the equipment of educational institutions with powerful computing devices. This reduces the cost of it's implementation.

Ji-Seong Jeong, Mihye Kim, and Kwan-Hee Yoo offer quite interesting developments [5]. The proposed system provides the delivery and exchange of various applications of educational content by integrating a number of functions necessary for the deployment of a learning media service environment in the cloud. Due to this, teachers and lecturers will be able to create various forms of educational content, including text, images, videos, 3D objects and virtual reality-based (VR) and augmented reality (AR) virtual scenes using an author's tool, driven cloud in a common format. The authors offers to use them to develop the learning environment not only in schools but also in universities and institutes. The system may be adapted to the needs of each student / pupil, with the analysis of their benefits to the perception of new material, learning styles and content usage templates.

Research [1] shows that the augmented reality can be used to study, motivate and concentrate attention, in addition, interaction with the objects of the augmented reality improves the understanding and memorization. By creating an image of a 3D object, using the augmented reality, the teacher invites and attracts the attention of students, which is difficult to hold so long by means of other tools of learning (for example, traditional). In addition to these benefits, by involving augmented reality it will be possible to overcome the problems, repair and maintain of experimental equipment, students will be able to experiment without affecting the environment. Students are engaged with the augmented reality aspect and the opportunity to study the subject matter in the natural environment. Owing to augmented reality it becomes possible to make the virtual objects and the real world objects coexist, representing the object of a study, such as atomic structure that can be studied in a convenient representation, expanding the boundaries of space and time.

Augmented reality can be performed by means of [3]: marker position; geolocation; QR-code.

In the first case, the process is to associate 3D images, videos, or animations by means of the print mark of specific software. When a token passes through a webcam, the virtual layer contained in this token will be activated. As a result, if the marker perspective has changed, the virtual objects will change their orientation, and this will allow us to observe them in three-dimensional mode. Software programs such as Augmentaty, BlippAR, BuildAR and ARSights can be used for implementation in educational environments; these programs are intuitive and easy to use [15].

The second case is limited to the geolocation augmented reality, its purpose is to integrate augmented reality technologies, GPS, visual search system (CVS) and mapping (SLAM) [10; 11]. Such programs offer for users a structure for interacting with the city system, depending on their location at a certain point. By using the camera of their mobile device, the users my receive a physical image of the place and overlay the virtual layers of information that shows them. Users have a wide range of real-time data about upcoming events, history of the environment, institutions, etc. [12]

With regard to the third and last case, the use of augmented reality using QR codes, the interaction is perceived through a two-dimensional square form of the code that allows to store a large variety of alphanumeric combinations, which can then be visualized with a QR-reader installed on the mobile device; and it is through these codes that you can submit information [2].

Marcus Specht, Stefaan Ternier, and Wolfgang Greller identified the following technical components for mobile systems to work correctly with augmented reality [14]:

— flexible display systems, including system displays on the head, telephone camera and portable projectors. Media technologies are becoming more flexible and more profitable to produce. These technologies allow expanding the visibility of mobile users;
— sensor systems in mobile devices such as gyroscopes, GPS, electronic compasses, cameras, etc., microphones, and local area tracking systems;
— protocols and standards for the wireless network that support internal and external setup. They also provide access to multiplayer interaction in real-time augmented reality;
— mobile phones with computing power, allowing to visualize 3D objects and overlay in real time on an autonomous device;
— label tracking technology with six degrees of freedom, multi marker tracking and

hybrid tracking system. This technology is also associated with one of the most studied areas of augmented reality;
- the binding of augmented reality information based on location, based on textbooks, and storytelling. There are examples where augmented reality is used to support learning for this augmented reality experience is to be associated with educational programs, projects, or at least with the structure of objectives and approaches;
- flexible augmented reality browsers based layer with integration with social networks. Basically, the augmented reality system should be based on existing information channels and can provide information to users as a new kind of user interface. Therefore, the realization of mobile virtual reality for learning should consist of open interfaces to existing content and services.

Consequently, the inclusion of augmented reality to create new learning situations is allowed with the use of several principles, such as: the design of the environment that is flexible enough to provide the introduction of augmented reality components avoiding the technical problems; great attention should be paid to educational and didactic issues; essential elaboration of educational content to achieve a high level of mastering material beyond the mere reproduction for the learners and to allow teachers, as well as students to develop their IT competence; explore the methods that can be used in training with the elements of augmented reality; development of adaptive materials that can be used in different formats; training of pedagogical staff that can incorporate the added reality into educational practices and use it.